\documentstyle[multicol,aps,epsf]{revtex}
\begin{document}
\newcommand{\ii}{\'{\i}}
\draft
\title{Strange nonchaotic attractor in a dynamical system
 under  periodic forcing}
\author{Andr\'e S. Cassol$^{1}$, 
F\'abio L. S. Veiga $^{1}$ \cite{veiga}  and
Marcelo H. R. Tragtenberg$^{1,2}$  \cite{eu-email} \cite{absence}.} 

\address{$^{1}$ Depto. de F\ii sica, Universidade Federal de
Santa Catarina, Florianop\'olis, SC, Brazil, CEP 88040-900}
\address{$^{2}$ Theoretical Physics,
Oxford University, 1 Keble Road, Oxford OX1 3NP, England}

\date{\today}
\maketitle

\begin{abstract}
 We observe the occurrence of a strange nonchaotic attractor
in a periodically driven two-dimensional map,
formerly proposed as a neuron model and a sequence generator.
 We characterize this attractor through the study of
the  Lyapunov exponents, fractal dimension, autocorrelation
function and power spectrum. The strange nonchaotic attractor
in this model is a typical behavior, occupying
 a finite range of the parameter space.

\end{abstract}
\begin{multicols}{2}

\section{INTRODUCTION}

The strange nonchaotic attractor (SNA) is an object that has
chaotic attractor features like fractal dimension and
nondifferentiability (strangeness) but no exponential
sensitivity to initial conditions, i.e., its largest Lyapunov
exponent is nonpositive \cite{kapitaniak1}.  It has some analogies
with trajectories that have been found
in a study of   the  Frenkel-Kontorova 
model \cite{Bak} and of the Chirikov-Taylor map \cite{chirikov}. 
Aubry has shown \cite{aubryfk} that the
incommensurate ground-states of the Frenkel-Kontorova
model can undergo a {\em breaking of analiticity transition},
between a smooth and a fractal set. Shenker and Kadanoff
 \cite{kadanoff} calculated the fractal power spectrum
of  the fractal trajectory that
appears in the Chirikov-Taylor  map after the breakup of a KAM-like
surface a fractal power spectrum.
This map can be related to an incommensurate driving of
a nonlinear oscillator. Another example is the
accumulation point of the period-doubling cascade of the
logistic map \cite{feigenbaum}. However, this attractor occurs
in a zero measure  set in the parameter space. SNA as a typical
behavior, a finite measure set in the parameter space of a model,
 has been found 
in the context of nonlinear quasiperiodic external forcing
(i.e., the forcing of a signal with two incommensurate
frequencies) \cite{grebogi1}. 
The study of the SNA
has also been recently connected with the localization problem
 \cite{bondeson,ketoja}.

 A lot of work has been done in order  to characterize
the features of a SNA; its route of formation, autocorrelation
function  and power
spectrum.  Besides a period-doubling cascade, many different
routes for the formation of an 
SNA have been proposed: (1) the collision between a
period-doubled torus and its unstable parent torus  \cite{HH}; 
(2) the progressive fractalization of a two-dimensional ergodic
torus \cite{anish1} and, (3) for systems with quasiperiodic tori in
symmetric invariant subspaces, the loss of the transverse stability of
 a torus \cite{yacinkaya}. Another particular feature of
the SNA is that its  autocorrelation function does not decay 
with the time delay like that of the chaotic attrator. It can 
either be 
fractal or similar to the quasiperiodic case \cite{pikovsky1}.
 The power spectrum of a SNA  can be  singular continuous and
 in this case has fractal features as discussed in many
papers \cite{pikovsky1,romeiras,kuznet1,kuznet2,feudel2,anish2}

The aim of this paper is to answer the question: are SNAs
restricted to quasiperiodically driven nonlinear systems? We
believe that the answer is no. We studied a map (hereafter
called YOS map) that has been
proposed in the magnetic context to describe the behavior of 
an analog of the ANNNI model on the Bethe lattice
\cite{yos,ty} and which shows, for external periodic forcing,
the occurrence of a SNA \cite{cassol}.
 The same map exhibits typical
features of a neuron (like activation threshold, nerve blocking and
{\em rebound} behavior) and has been coined as a dynamical
perceptron of second order (related to the dimension two of the map),
 because it corresponds to a two-layer recurrent neural
network \cite{kt}. Independently, Kanter {\em et al.}
\cite{seqgen} proposed a recurrence relation scheme known as
  a {\em sequence generator}, which is  essentially the same map,
generalized for any number of dimensions. This map can also be viewed
as a discrete-time, nonlinear oscillator, for some values of the 
parameters.

Anishchenko {\em et al.} \cite{anish2} have claimed that they
found a SNA through periodic  driving of a map, but Pikovsky {\em et al.}
\cite{pikovsky2} have shown that it was a chaotic
attractor with a tiny Lyapunov exponent. We show
 that Anishchenko {\em et al.}'s attractor is indeed
strange chaotic, but ours is strange nonchaotic and related to 
 quasiperiodic attractors.

This paper is organized as follows. Section II is dedicated to
describing the map and its  attractors, mainly the strange
nonchaotic one. In section III we characterize this attractor
through its Lyapunov exponents, fractal dimension,
autocorrelation function and power spectrum. The conclusions are
addressed to section IV.

\section{THE MAP AND ITS ATTRACTORS}

The two-dimensional YOS map which this paper is concerned with is given by

\begin{eqnarray}
\label{mapyos}
x_{n+1} & = & \tanh \biggl [ \frac {x_n - \kappa y_n + H(n)}{T}
\biggr] \\ y_{n+1} & = & x_n \nonumber
\end{eqnarray}

The YOS map was initially proposed \cite{yos} to model the
mean magnetizations $(x_n,y_n)$ of the $(n^{th};(n-1)^{th})$-shells 
of a Bethe lattice,
for an  analog of the ANNNI model \cite{selke1,julia1,selke2} in
a constant magnetic field (H was independent of n). Here,
we extend it for a nonuniform field.  T is the
temperature and $\kappa = -J_2/J_1$ where $J_1 (J_2)$ is the exchange 
coupling between nearest (next-nearest) neighbor spins on a Bethe lattice. 
Yokoi {\em et al.} obtained the phase diagram
of this model at zero field \cite{yos}. Tragtenberg and
Yokoi studied the effect of finite uniform field \cite{ty}. 
 A sinusoidal wave $H(n)  = H_0 \cos(2 \pi \omega n)$ is the 
particular form of $H(n)$  we will adopt throughout this paper,
 representing a shell-dependent external field/input. 

Kinouchi and Tragtenberg \cite{kt} 
studied the properties of the map as a
neuron model, where $x_n$ is the action potential of the neuron
at time $n$.  $1/T$  and $-\kappa/T$ are the weight factors for
the two previous states of the neuron. $H(n)/T$ is
naturally defined as the external current as a function of the
discrete time $n$.  They showed that the map exhibits many
neural features.

Kanter {\em et al.}\cite{seqgen}
 proposed the following real number sequence generator:

\begin{equation}
s_l = \tanh \biggl[ \beta \sum_{n=1}^{N} W_{n} \; \;  s_{l-n}
\biggr]
\label{sg}
\end{equation}

\noindent and studied this map in the context of time-series and neural
networks. $ H(n)$ could be introduced for representing
 a time dependent input signal.

\begin{figure} 
 \narrowtext 
 \centerline{\epsfxsize=3.1in 
 \epsffile{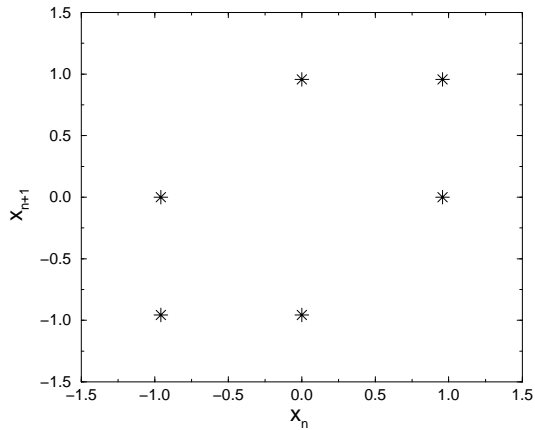}}
 \vskip 0.1true cm 
 \caption{ First return map for the cycle of period 6 of the YOS map
 with $\kappa =$ 1 , $T = $ 0.5 and $H_0 =$ 0. } 
 \label{1-6}
 \end{figure}

\begin{figure} 
 \narrowtext 
 \centerline{\epsfxsize=3.1in
 \epsffile{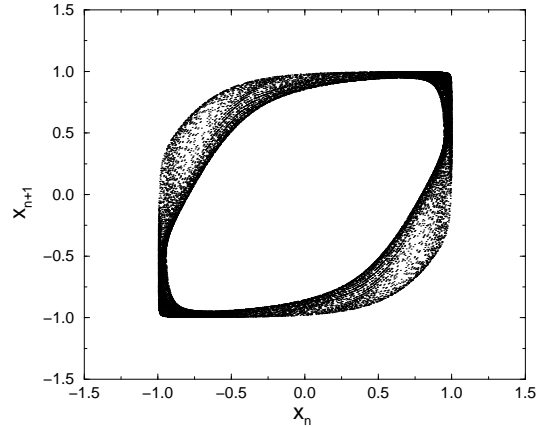}} 
 \vskip 0.1true cm 
 \caption{ First return map of the strange nonchaotic attractor
 of   YOS map with $\kappa = 1 , T = 0.5, H_0 = 0.17  $ and $\omega =
 0.14$. The initial condition is $(x_0,y_0)=(1,1)$. Here are
 represented 30 000 iterations, after neglecting the first 10,000.}  
 \label{sna} 
 \end{figure}

From the purely dynamical system point of view, the YOS map
 represents a nonlinear oscillator for the  range  of the
parameters considered in this paper, and $H(n)$
represents a time-dependent external input.
The attractors of map YOS can be fixed points, Q-cycles (cycles
of period Q), quasiperiodic, chaotic or strange nonchaotic
 \cite{yos,ty,cassol,fabio}.
For the parameters $\kappa = 1, T = 0.5, H_0 = 0  $ the system
oscillates with period 6 (see Fig. \ref{1-6}). But, for  different
 values of $H_0$ and $\omega$, 
keeping the same $\kappa $ and $T $, the attractor becomes
richer (see Fig. \ref{sna}).

This kind of  attractor can also be obtained for other values of
 $\omega$, and is therefore  characteristic of this periodically
driven nonlinear map.  For others examples of SNA 
of this map see \cite{fabio}.

\section{CHARACTERIZATION OF THE SNA}

A SNA is a fractal object with no exponential sensitivity to
initial conditions (the SNA at the accumulation point of the
period-doubling bifurcations of the logistic map has null
Lyapunov exponent but has  polynomial
sensitivity to initial conditions \cite{tsallis}). In order to 
characterize this attractor, we investigated the largest
Lyapunov exponent, fractal dimension, autocorrelation function and
power spectrum. 

\subsection{Lyapunov exponents: na\"{\i}ve and more accurate
calculation}

Before calculating the Lyapunov exponent of the attractor of
Fig. \ref{sna}, let us briefly discuss the sensitivity to
initial conditions of one of the  attractors
 studied by Anishchenko {\em et al.} \cite{anish2}.

They proposed a four-dimensional map made up by two asymmetrically
coupled circle maps, with two coupling parameters (A and $\gamma_2$). They
argued that the for $\gamma_2 =0$ the system (1) can be a circle map with
quasiperiodic forcing. A small value of $\gamma_2$ will make it an autonomous
four-dimensional map that could show a SNA.

	This  four-dimensional map is given by

\begin{eqnarray}
\label{mapanish}
x_{n+1} & = & x_n + \Omega_1 - \frac {K_1}{2 \pi} \sin (2 \pi
x_n) +
\gamma_1 y_n \nonumber \\ 
        &   &  + A \cos(2 \pi u_n) \; \;  \bmod \; 1, \nonumber
\\ y_{n+1} & = & \gamma_1 y_n - \frac{K_1}{2 \pi} \sin(2 \pi
x_n),  \\ u_{n+1} & = &  u_n + \Omega_2 - \frac{K_2}{2 \pi} \sin
(2 \pi u_n) +
\gamma_2 (y_n + v_n) \bmod 1, \nonumber \\
v_{n+1} & = & \gamma_2 (y_n + v_n) - \frac {K_2}{2 \pi} \ sin(2
\pi u_n) .  \nonumber
\end{eqnarray}

For the set of parameters $\Omega_1 = 0.5, \Omega_2 =
(\sqrt{5}-1)/2, K_2 = 0.03,  A=0.4, \gamma_1 = \gamma_2 = 0.01 $
and $ K_1 = 0.8784 $,  Anishchenko {\em et al.} 
claimed the attractor is strange
nonchaotic. They found a null largest Lyapunov exponent within the
numerical accuracy of the method they used.

A positive definite largest Lyapunov exponent corresponds to an
exponential expansion of an hypercube
 of initial conditions in at least one of
the directions  of the phase space. In other terms, we can study
the sign of the largest Lyapunov exponent of a map by studying the
stretching and contraction of a hypercube of initial conditions. Here
we present a simpler version of this procedure, studying only the
evolution of the distance between trajectories generated by only
two initial conditions. 

\begin{figure}  
 \narrowtext 
 \centerline{\epsfxsize=3.1in
 \epsffile{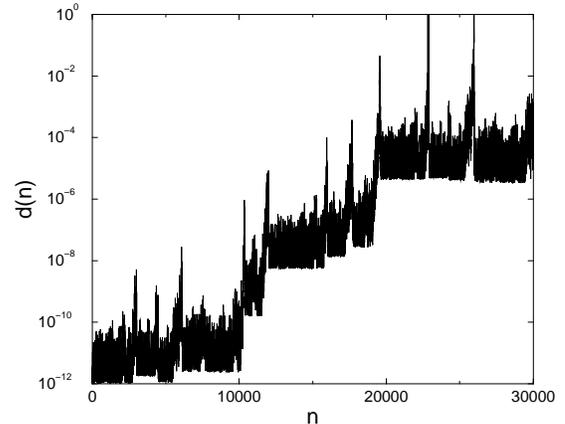}} 
 \vskip 0.1true cm 
 \caption{ Distance between attractors with different
 initial conditions $d(n)$ as a function
  of the number of iterations $n$, for the map
    (\ref{mapanish}) for the parameters $\Omega_1 = 0.5,
  \Omega_2 = \frac{\sqrt{5}-1}{2}, K_2 = 0.03, A=0.4,
  \gamma_1 = \gamma_2 = 0.01 $ and $ K_1 = 0.8784 $.
  The two set of
 initial conditions are $x_0= y_0= u_0= v_0=0.7$
  and $x_0'=y_0'=u_0'= v_0'= (0.7-10^{-12})$. The
 largest Lyapunov exponent is   $\lambda_+ = (0.8 \pm 0.3)  10^{ -3 } $.  }   
 \label{Lyapanish}
 \end{figure}

We take two different sets of initial values of (x,y,u,v) and
calculate the distance $d(n)$ between the trajectories  generated by
each set,  as the number of iterations is increased.
That is perhaps the na\"{\i}vest way to
investigate the largest Lyapunov exponent of a map.  The first
set we take as  $x_0 = y_0 = u_0 = v_0 = 0.7$ and the second as
$x_0' = y_0' = u_0' = v_0' = (0.7-10^{-12})$. Figure
\ref{Lyapanish} represents the first 30 000 iterations of the 
evolution of d(n).  

We can see
at first sight that the system is chaotic, since the distance
between the trajectories with different 
initial conditions grows exponentially.  A simple
estimation of the slope of the rugged curve leads to
 $(0.8 \pm 0.3)  10^{-3}$ for the largest Lyapunov exponent 
$\lambda_+$, where
we have assumed that the behavior of the distance is 
governed by this exponent and given by
\begin{equation}
\label{naiveLyap}
d(n) \approx  d(0) \exp (\lambda_+ n).
\end{equation}
This result agrees with surprising accuracy with
that obtained by Pikovsky and Feudel \cite{pikovsky2}, using
the Wolf-Swift-Swinney-Vastano algorithm \cite{wolf}.

Then, this na\"{\i}ve method of checking the sensitivity of
initial conditions seems to be powerful and we will use it as
well as the more accurate method due to
Eckmann-Kamphorst-Ruelle-Ciliberto (EKRC) \cite{eckmann} to calculate
the largest Lyapunov exponent of the attractor of Fig. \ref{sna}.

\begin{figure} 
 \narrowtext 
 \centerline{\epsfxsize=3.1in
 \epsffile{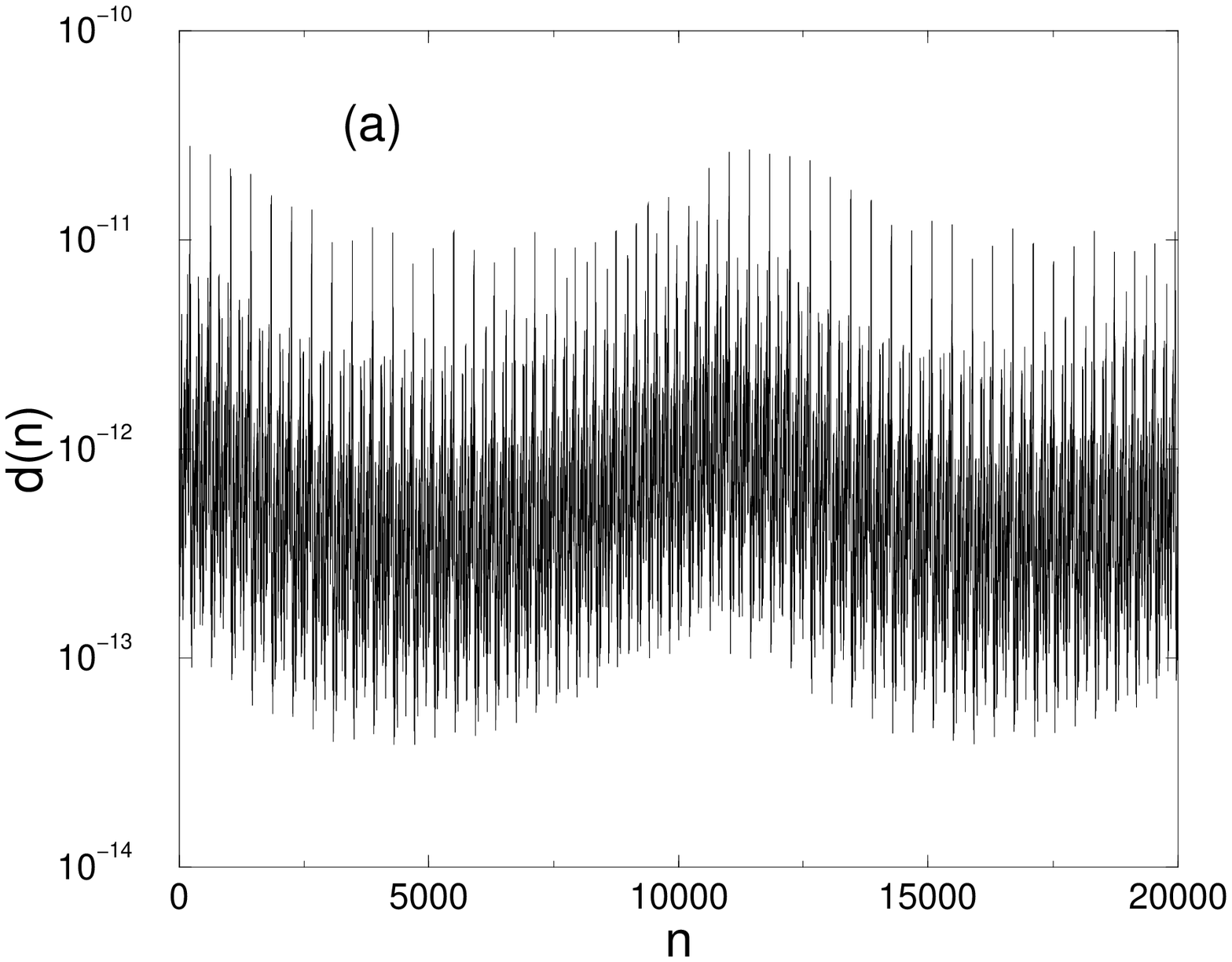}}
 \centerline{\epsfxsize=3.1in
 \epsffile{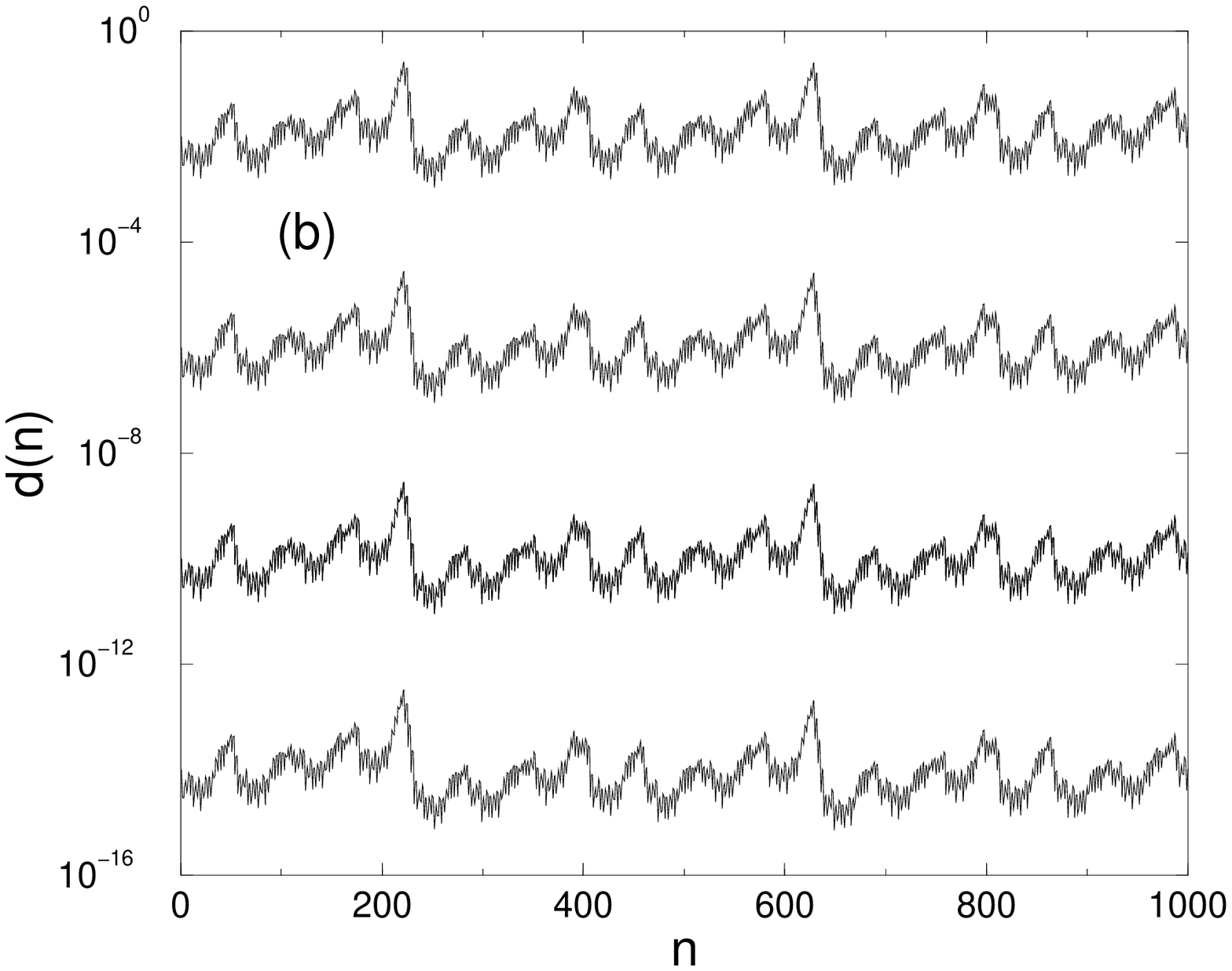}} 
 \vskip 0.1true cm 
 \caption{Distance between trajectories  with different  initial
  conditions  $d(n)$ as a
 function of the number of iterations $n$ for the SNA of the
 Fig. \ref{sna}, when:
 (a) one initial condition is $x_0 = y_0 = 1$ and the other is
 $x_0'=y_0'=(1- 10^{-12}) $. Notice the self-similar like
  structure of the peaks;
 (b) one initial condition is ever taken as  $x_0 = y_0 = 1$ and the other
 has the form $x_0' = y_0' = (1-\Delta) $, for  $\Delta = 10^{-2}, 10^{-6},
 10^{-10}$ and $10^{-14}$. The distances between the trajectories
  remain limited even in the limit of large number of iterations, 
  pointing out to  a null largest Lyapunov exponent, and exhibiting
 a scale invariance.  }
 \label{naivesna} 
 \end{figure}

Fig. \ref{naivesna}a exhibits some self-similarity in the behavior
of $d(n)$ as a function of $n$.
 We took many pairs of initial conditions such that the distances
between  the initial conditions from each pair were $10^{-2}, 10^{-6},
10^{-10} and 10^{-14}$. Then we calculated
how $d(n)$ vary for each pair as a function
of the iterations, for  the  SNA of the Fig. \ref{sna}. The result 
 is represented in Fig. \ref{naivesna}b. The various curves d(n) x n
 have a scale invariant like
behavior, i.e., for various values of the difference
in initial conditions the evolution with the iterations is rather
similar, preserving the form for different scales  of distance.
 Moreover, none of the curves show exponential
divergence, although they may vary within few orders of
magnitude. It suggests a null largest Lyapunov
exponent. We confirmed this using  calculations
based on the EKRC method shown below.  This
behavior (scale invariance in distance and zero largest Lyapunov
exponent) is  similar to that of a typical quasiperiodic
attractor.

\begin{figure} 
 \narrowtext 
 \centerline{\epsfxsize=3.1in
 \epsffile{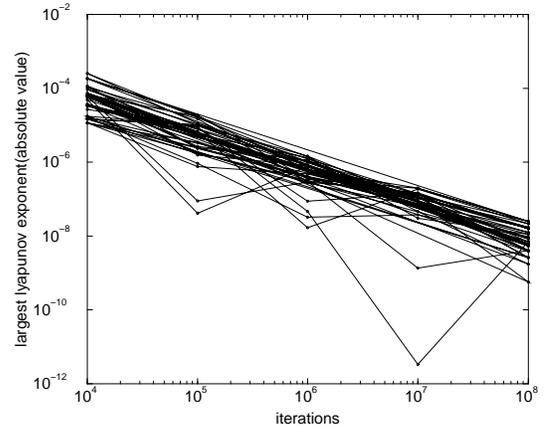}} 
 \vskip 0.1true cm 
 \caption{ Plot of the absolute values of the
 largest Lyapunov exponent approximants for the  SNA of
 Fig. \ref{sna} as a function of the number of iterations $n$ (the first
 $10^4$ were neglected). The initial conditions 
 considered were ($x_0,y_0$) = ($\pm 1, \pm 0.5, 0 ; \pm 1, \pm
 0.5, 0$).  The absolute value of the approximants scales as
   $  | \lambda_{+}| \sim {n}^{-1}$, indicating a zero largest
 Lyapunov exponent. The method used here is due to
 Eckmann-Kamphorst-Ruelle-Ciliberto [31] .}
 \label{lyapsna} 
 \end{figure}

Fig. \ref{lyapsna} shows  the behavior of the absolute value of
the largest Lyapunov exponent approximants as a function of the number
of iterations (neglecting the first 10,000),  
for the attractor of the Fig. \ref{sna}. This attractor has the same
shape for many initial
conditions: we took ($x_0,y_0$) = ($ \pm 1, \pm 0.5, 0 ;\pm 1, \pm
0.5, 0$). The calculations were performed using the
 Eckmann-Kamphorst-Ruelle-Ciliberto
method.  They do  indicate that the 
 largest Lyapunov exponent is zero, since 
 the  absolute values of its approximants scale with the number of
iterations $n$ as $ | \lambda_{+}| \sim   n^{-1}$. Using the
 same method, we
found $\lambda_{-}=-0.709971 \pm 0.000001$, for the smallest Lyapunov exponent.

\subsection{Fractal dimension}

\begin{figure} 
 \narrowtext 
 \centerline{\epsfxsize=3.1in
 \epsffile{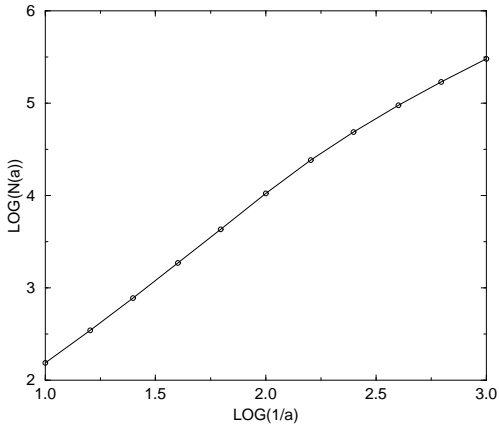}} 
 \vskip 0.1true cm 
 \caption{ Number of visited boxes $N(a)$ as a function
 of the length of the box edge  $a$, for the attractor of the 
 Fig. \ref{sna}. The first $10^4$ iterations were neglected and the
 next $10^9$ considered.}
 \label{nepsilon} 
 \end{figure}

The SNA of the Fig. \ref{sna} is a complex geometrical object
with a fractal Hausdorff dimension ($D_F$). In order to find out this
dimension we have used
the box counting method \cite{boxcount}. The diagram with
 the number of  boxes $N(a)$ visited by the SNA of Fig. \ref{sna}
 as a function of the edge
length $a$ is shown in Fig.\ref{nepsilon}. 
 The initial condition is $(x_0,y_0)=(1,1)$, and the first $10^4$
iterations were discarded. We considered the next $10^9$ iterations,
and box edges between $a=10^{-1}$ and $a=10^{-3}$. Even with this
number of iterations,   
 we can see that box edges smaller than $10^{-2.6}$ lead to 
artificially small values of the 
fractal dimension. However, a larger number of iterations
were computationally prohibitive. The 
same behavior was  observed in \cite{ding}. 

\begin{figure} 
 \narrowtext 
 \centerline{\epsfxsize=3.1in
 \epsffile{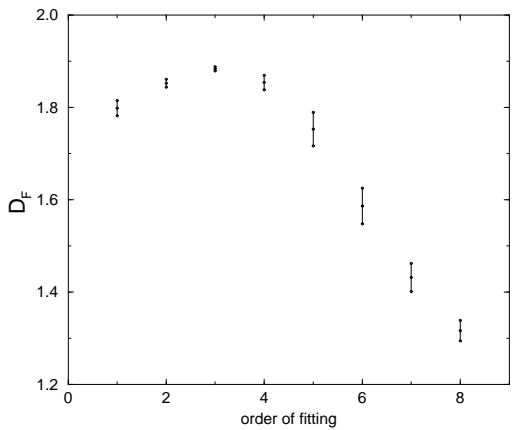}} 
 \vskip 0.1true cm 
 \caption{ Hausdorff dimension $D_F$ of the attractor of Fig. \ref{sna}
 as a function of the order {\em i}of fitting, for 
   sets of four consecutive points of
 the Fig. \ref{nepsilon}. The order or fitting {\em i } denotes that the
 slope (and error bars) were calculated for the 
 ${(i,i+1,i+2,i+3)}^{ths}$ points of Fig. \ref{nepsilon} (the counting
 order begins
 from the left to the right). $D_F$ is the value
 of the slope of the line determined by these point sets through least 
 squares fitting.}
 \label{fractald} 
 \end{figure}

Fig. \ref{fractald} has been constructed
by taking ordered sets of four consecutive points 
of  Fig. \ref{nepsilon} and
calculating the linear coefficient of the best straight line determined
by them. Error bars follow from least squares fitting.
The leftmost point of this figure represents the fitting of the
four smallest values of log(1/a). The point of order 2 represents the
fitting of the second to fifth point of Fig. 6, counting from the left
to the right, and so on.
 Then, we conclude that the fractal
dimension is $D_F =1.80 \pm 0.09$. This is 
the same value found in Ref.  \cite{ding} for the attractor
of Grebogi {\em et al.} \cite{grebogi1}, but they considered this
value quite uncertain (since it was obtained from just three points).
 In the same reference, Ding {\em et
al.} used heuristic arguments to conjecture
 that $D_F = 2$ and found no contradiction with the result they
 numerically found. But this value is definitely far from
the value we obtained for the attractor of the Fig. \ref{sna}.
  The evidence points to the fractal character of this object.

\subsection{Autocorrelation function}

The normalized  autocorrelation function of an attractor \{ $x_n$\}
 can be defined as

\begin{equation}
C(\tau) = \frac{ \sum_{n=1}^N  x(n)x(n+\tau)} 
{\sum_{n=1}^{N} {x^2(n)} }.
\end{equation}

The calculation of the autocorrelation function can 
give clues about the nature of the attractor in question.
Its fractal character can indicate the fractality of the attractor.
 However, when
 we are in the presence of a SNA we can observe at least two kinds of
autocorrelation function: fractal or quasiperiodic \cite{pikovsky1}.

\begin{figure} 
 \narrowtext 
 \centerline{\epsfxsize=3.1in  
 \epsffile{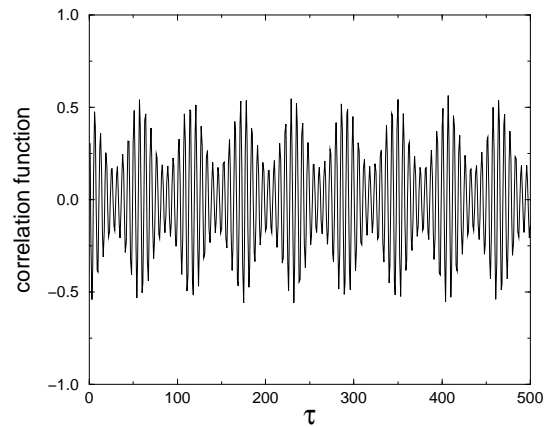}} 
 \vskip 0.1true cm 
 \caption{ Autocorrelation as a function of the time delay $\tau$
  for the attractor of the
 Fig. \ref{sna}, which has the same form of the
 autocorrelation function of a quasiperiodic attractor.
  Here, we discarded the first
 $10^4$ iterations and considered the next $10^5$ iterations.  }
 \label{corr} 
 \end{figure}

Fig. \ref{corr} represents the autocorrelation function of the
attractor of the Fig. \ref{sna}, and is very similar to those related
to quasiperiodic attractors, like that found in Ref.
\cite{pikovsky1} for the strange nonchaotic attractor of the model C
 defined therein. We
neglected the first $10^4$iterations and considered averages over
the next $10^5$ iterations.

\subsection{Power spectrum}

The first step in investigating the power spectrum of an attractor given
by a sequence \{$ x_n$\} is defining its discrete Fourier transform
\begin{equation}
s(w,N) = N^{-1/2} \sum_{n=1}^{N} x_n e^{i 2 \pi w n}.
\end{equation}
Then,we can define the power spectrum of the attractor as:

\begin{equation}
P(w) = \lim_{N \rightarrow \infty} < {|s(w,N)|}^2>.
\end{equation}

The power spectrum of periodic attractors consists of $\delta$-peaks
at the harmonics of the fundamental frequency, whilst in the chaotic
case the spectrum is continous. For a quasiperiodic case characterized
by two incommensurate frequencies $\omega_1$ and $\omega_2$, the
spectrum contains all the frequencies of the form $n \omega_1 +  m
\omega_2$.

Many works report power spectra of a SNA with 
singular continuous \cite{kadanoff,pikovsky1,kuznet1,feudel2}
character,   like those found in some models of quasiperiodic
lattices and quasiperiodically forced quantum systems
\cite{aubry1,godreche,aubry2,luck}. This spectrum has a fractal appearance,
where there are peaks weaker than $\delta$-functions distributed along
a self-similar landscape.

\begin{figure} 
 \narrowtext 
 \centerline{\epsfxsize=3.1in 
 \epsffile{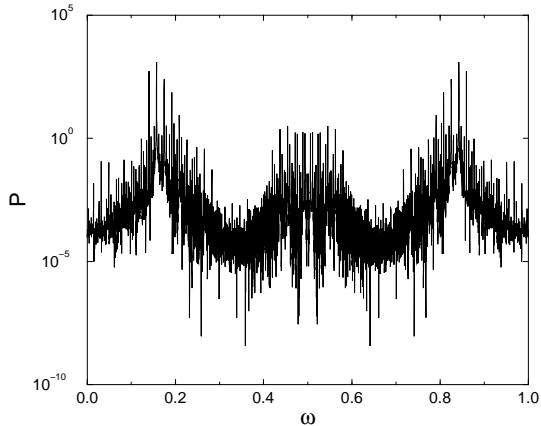}} 
 \vskip 0.1true cm 
 \caption{ Power spectrum of the attractor of Fig. \ref{sna}, with a
 fractal appearance. The first $10^4$ iterations were discarded and
 the following $10^4$ considered. Notice the resonance
 at the driving frequency  $\omega = 0.14$.}
 \label{powerspec} 
 \end{figure}

Fig. \ref{powerspec} shows the power spectrum of the SNA of the
 Fig. \ref{sna}. It  has  many scales of peaks, exhibiting a fractal
appearance. We neglected the transient of the first $10^4$  iterations
and took the next $10^4$. The detailed study of the fractal character
 of  this  power spectrum as well as
a renormalization group approach for it  is the subject of a
forthcoming publication.

\section{Conclusions}

We are able to show that a strange nonchaotic attrator can
result from the dynamics of a periodically driven 
 nonlinear oscillator,
 the YOS map. We have shown that a fractal object with
zero largest Lyapunov exponent can emerge from this dynamics
in a finite range of the parameters space. The Hausdorff dimension
 of this object is $ D_F = 1.80 \pm 0.09$.  Its correlation
function oscillates like those of quasiperiodic attractors
and its power spectrum has a fractal (or multifractal) appearance.

\section{acknowledgements} MT acknowledges O. Kinouchi by suggesting that
this attractor could be strange nonchaotic, is grateful to N.N.Oiwa for
valuable discussions, mainly for his criticism against the
character of the attractor reported here, thanks C. Denniston for
bringing to our attention Ref.  \cite{kadanoff}, and acknowledges
FINEP/Brazil for the partial financial support. We acknowledge
J.M. Yeomans for the careful reading of the manuscript. A.S. Cassol and
F.L.S. Veiga  acknowledge CAPES/CNPq and CNPq, respectively, for
partial financial support.

\end{multicols}
\end{document}